\documentstyle[11pt,twoside]{article}
\newcommand{\Acknow}[1]{\par\vspace{5mm}{\bf Acknowledgements.} #1}
\begin{document}
\begin{center}
{\large\bf Optical Gravitational Lensing Experiment.\\
\large\bf OGLE-2 -- the Second Phase of the OGLE Project}
\vskip1.3cm
{\large A.~~U~d~a~l~s~k~i, ~~M.~~K~u~b~i~a~k ~~and~~M.~~ 
S~z~y~m~a~{\'n}~s~k~i}
\vskip5mm
{Warsaw University Observatory, Al.~Ujazdowskie~4, 00-478~Warszawa, Poland\\
e-mail: (udalski,mk,msz)@sirius.astrouw.edu.pl}
\vskip13mm
{\it ~}
\end{center}
\vskip1.5cm

{\footnotesize\begin{center}{ABSTRACT}\end{center}
We describe a new 1.3 m Warsaw telescope located at Las Campanas 
Observatory and new instruments of the second phase of the Optical 
Gravitational Lensing Experiment -- OGLE-2. Results of first observations are 
also presented.}
\vskip1cm

\vspace*{12pt}
{\bf 1. Introduction}
\vspace*{9pt}

The Optical Gravitational Lensing Experiment (OGLE) project is a long term 
project with the main goal of searching for the dark matter with microlensing 
phenomena. The first phase of the project (OGLE-1) started in 1992 and 
observations were continued for four consecutive observing seasons through 
1995. The \hbox{1-m~Swope} telescope at the Las Campanas Observatory, Chile, operated by 
Carnegie Institution of Washington, with ${2048\times2048}$ Ford/Loral CCD 
camera were used during the entire program. 

The project was very successful, the most important discoveries
included:\\
-- discovery of the first microlensing events toward the Galactic bulge
(Udalski {\it et al.} 1993),\\
-- first determination of the optical depth for microlensing toward
the Galactic bulge (Udalski {\it et al.} 1994a),\\
-- discovery of the first microlensing by a binary object -- OGLE-BUL-7
(Udalski {\it et al.} 1994b),\\
-- first system of detection of microlensing events in progress
(Early Warning System, EWS, Udalski {\it et al.} 1994c).

In addition to the microlensing search, many side projects were
conducted using  the huge database of photometric measurements. The most
important included:\\
--  the Galactic structure analysis -- evidence for existence of a bar
in the Galaxy (Stanek {\it et al.} 1996),\\
--  variable stars search in the direction of the Galactic center:
"Catalog of Periodic Variable Stars in the Galactic Center" (Udalski
{\it et al.} 1997) including 2861 objects,\\
--  photometry of globular clusters -- discovery  of numerous detached
eclipsing binaries in $\omega$~Cen and 47~Tuc (Ka{\l}u{\.z}ny {\it et al.} 1996),\\
-- photometry of dwarf galaxies in Sculptor (Ka{\l}u{\.z}ny {\it et al.} 1994) and
Sagittarius (Mateo {\it et al.} 1996).

Unfortunately the project suffered from many limitations with the most
severe  -- limited availability of the telescope time. Therefore the
observations were  limited to the Galactic bulge, and the covered area
on the sky was relatively  small. It was clear from the very beginning
that OGLE-1 is only the pilot  phase of the project and a
substantial upgrade --  a dedicated telescope and new  generation 
instruments -- are necessary to accomplish the main goals of the OGLE 
project. 

In this paper we describe new hardware, software and instrumentation of
the OGLE-2 -- the second phase of the OGLE project which  started in 1997.
In Section~2 we  describe briefly the history of project, and the
following Sections cover details  of the telescope, buildings, CCD
camera and other new instruments. Brief description  of the computer
equipment and software is also provided. Finally, first results  of tests
and observations are presented. 

\vspace*{12pt}
{\bf 2. History of the Project}
\vspace*{9pt}

In the late 1991 the funds  for the OGLE microlensing project were granted by 
the State Committee for Scientific Research to the team of astronomers from 
the Warsaw  University Observatory. One of the goals of the grant was to 
design and build a  1-m class telescope dedicated for massive photometric 
surveys of dense stellar  fields. 

After careful inspection of the market the DFM Engineering Inc.
(Longmont,  Colorado, USA) was selected as a manufacturer of the
telescope. The company   offered 1-m class telescope at reasonable price
and fulfilling requested  parameters. The contract was signed on
November 24, 1992. 

The good location of the telescope was the next crucial point. Long
collaboration of Polish astronomers  with astronomers from Carnegie
Institution of Washington, excellent climate  condition and visibility
of Magellanic Clouds and Galactic center  made the  Las Campanas
Observatory, Chile, operated by Carnegie Institution of Washington  the
best choice. Thanks to a very friendly and cooperative attitude of 
Carnegie's astronomers the  "Memorandum of Understanding" was signed in
1995 between Carnegie Institution of Washington,  Princeton University
and Warsaw University Observatory  defining the terms of presence of the
1.3-m Warsaw Telescope on the mountain. 

First civil engineering works started in August 1995. In October 1995 the
telescope  enclosure manufactured in Poland and shipped to Chile was
assembled. At the same time  construction of the observer's house 
started. In January  1996 the dome was assembled and mounted atop of the
telescope  enclosure. A week later assembling of the telescope started. 

The "first light" of the telescope was collected on February 9, 1996.
The  telescope was ready for operation in the mid-February 1996 and the
next  step was assembling and testing the instruments. The first
"electronic light",  taken with the main CCD camera was registered on
July 18, 1996. To the  end of 1996 remaining instruments: autoguiding
system and  filter wheel were  commissioned. Also the software for
hardware control and data reduction was  finished. Regular observations
of the second phase of the OGLE project  started on January 6, 1997. 

Fig.~1 shows the telescope building and observer's house at Las Campanas
Observatory. 

\vspace*{12pt}
{\bf 3. Telescope}
\vspace*{9pt}

The main feature which should characterize the telescope designed for
massive  photometric survey is a large field of view with good image
quality. The  operation of the telescope should  be as automatic as
possible, allowing interfacing with an external computer. The mechanical
design  should be as precise as possible, with good tracking in both
axes. All these  features  can ensure good throughput and expandability
of the system in the  future. 

\newpage
Fig.~2 shows the 1.3-m Warsaw telescope with instruments.  It is a
Richey-Chreti{\'e}n system, with the primary mirror of diameter  1300~mm
and primary focal ratio 1:2.8. The effective  focal ratio of the  system
is 1:9.2 giving the focal scale of 17.4~arcsec/mm. The diameter of the
secondary mirror is 432~mm. Both mirrors are made from Corning ULE
(Ultra Low Expansion) substrate. The field of view of the bare system is
about 15~arcmin (80\% of energy in 0.5~arcsec disk). The optics  was
manufactured by Tinsley Lab.\ (Richmond Ca, USA). 

The full 1.5 degree field of view is obtained with the additional three 
element corrector made of the BK7 Schott glass. Thus, the  corrector
cuts off the far ultraviolet part of the optical spectrum. The image 
quality is better than 80\% of energy in 0.5~arcsec disk over the entire
field of  view. The focal plane is about 200~mm behind the last optical
surface of the  corrector.  All  optical surfaces were anti-reflective
coated with ${\rm MgF_2}$.

The optics are mounted in the equatorial fork mounting consisting of the
 pedestal, fork, truss optical tube with  primary mirror cell, and
secondary  mirror mounting with focus housing. Due to relatively short
focal length of  the primary, the  optical tube can be short, compact
and stiff which makes the  telescope very stable and well protected
against wind gusts.

The primary mirror support consists  of an airbag on which the mirror
sits,  four radial teflon hard points which hold approximately 10\% of
the weight of  the mirror and a set of radial supports which load is
dependent on the  position of the telescope. The load of the primary is
measured by three load  cells and  electronic circuit keeps this load
constant by changing the air  pressure in the bag. The air bag supports
approximately 90\% of the weight of  the primary mirror. Tests showed
that position of the primary mirror is kept  within 5~$\mu$m over the
entire range of useful telescope positions. 

The secondary mirror is mounted through the central hole to the
collimation  plate  with the invar plug. The collimation  plate is
attached to the focus  housing. Position of the collimation plate (and
the secondary mirror) can be  regulated with four collimation screws.
The distance of the secondary mirror  mounting relative to mirror cell
is held by four invar rods to minimize   changes with temperature.
Position of the secondary mirror relative to the  primary mirror can be
changed with the focusing mechanism driven by stepper  motor with the
resolution of about 2~$\mu$m. 

Tracking and slewing of the telescope are achieved with the friction
drives in  both RA and DEC axes and are driven by DC torque-limited
servo motors. This  ensures high acceleration and deceleration rates and
smooth operation of the  telescope. Position of the telescope is read
with incremental optical  encoders. Pointing accuracy of the telescope
is about 15~arcsec (rms) over the  entire sky (${z<70}$~deg). 

\vspace*{9pt}
{\bf 4. Telescope Building and Observer's House}
\vspace*{6pt}

The telescope is placed in the 11~m tall building (Fig.~1) consisting
of the  cylindric enclosure and 7.45~m dome manufactured by Ash-Dome
Inc.\  (Plainfield Il., USA). The enclosure was designed and
manufactured by  Mostostal-Project SA (Warsaw, Poland). To minimize
influence of the building  on the excellent seeing conditions of the Las
Campanas Observatory the  enclosure is a light steel construction
ensuring fast heat exchange and  thus the adaptation to the external
thermal conditions. There are only two floors in the building: ground
floor and telescope floor at 6-m level. For easy,  natural ventilation,
several louvers were mounted on both levels. Some of them  are equipped
with motors for computer controlled operation. At the ground  floor
there is a small control room where only the most necessary hardware for
operation of the telescope and instruments is stored:  telescope control
hardware, CCD power supplies and instrument control PCs. 

The telescope is operated from the small observer's house  separated by
about  15~m from the telescope building. The observer's house  is
insulated and equipped with forced ventilation to ensure minimum heat
exchange with the environment. The control  building  consists of the
large room where all computers, data storage  hardware, etc.\ are kept,
small kitchen, bathroom and storage room and provides the observer
with comfortable  working conditions. 

\vspace*{9pt}
{\bf 5. CCD Camera}
\vspace*{6pt}

The 1.3-m Warsaw telescope is dedicated for long-term photometric
surveys of the dense stellar fields. Therefore the CCD camera is the
main and the only  instrument attached to the telescope. This makes it
possible to achieve  extremely  stable system, better tuned for
photometric purposes than  regular, general purpose,  multi-instrument
telescopes. 

The necessity of covering large fields in the course of the photometric
surveys puts special constraints on the camera design. We decided to
design a modular system, easily upgradeable in the future as more and
more of the optical plane  will be covered by CCD detectors. As the
first step we designed and built  the first generation single chip
camera. It will be replaced in the future by  next generations of more
advanced mosaics with eight and more detectors. 

The first generation camera detector is a ${2048\times2048}$ pixels Scientific
Imaging Technologies (SITe) CCD chip (SITe424B) with pixel size of
24~$\mu$m  giving the scale of 0.417~arcsec/pixel at the focus of the
telescope. The chip  is cryogenically cooled with liquid nitrogen (LN2).
The cryostat holding the  detector is a ND-8 type dewar purchased from
Infrared Laboratories Inc.\  (Tucson, Az., USA). The standard design of
the dewar was slightly modified to  give larger room for the detector
and possibility of working in inverted  (look-up) position. 3.5~inch 
diameter quartz  window is attached to the lid plate of the dewar. The
window is anti-reflectively coated with ${\rm MgF_2}$  layer. The LN2
capacity of the dewar is 3.2 liter. 

The dewar is shown in Fig.~3. The CCD detector is mounted in the
four-arm spider-like mounting attached  rigidly to a removable dewar lid
(top plate). The bottom part of the "spider"  is made of brass and is
kept at low temperature. Four posts machined from G-10  fiberglass
isolate the bottom part of the "spider" from the ambient  temperature
top plate. The chip is held in a rectangular frame made of G-10 
fiberglass, which position relative to the lid can be adjusted by four
2--56  screws mounted in 0.005~in. wall stainless steel tubes isolating
frame from   the top plate. The chip is pushed to the fiberglass frame 
from the bottom by  a spring loaded, rectangular cold finger mounted on
the bottom part of the  "spider". Beside the stainless steel spring,
four copper straps connect the  cold finger with the bottom part of the
"spider". By changing  their width and  length one can regulate heat
exchange and minimum temperature of the cold finger (second temperature
reduction stage). Additionally, four resistors epoxied in the cold
finger serve as  heaters. Also temperature sensor is epoxied there.
Thermal contact between the chip and cold finger is improved by use  of
thermal conductive cryogenic grease. The top plate and the CCD mounting
form a single rigid element and can be easily removed from the rest of
the dewar after disconnecting cables  connecting dewar electronics with
hermetic 55-pin connector placed on the side  wall of the dewar. 

When the dewar lid is bolted to the dewar, the bottom part of the
"spider"  mounting of the CCD connects through the socket-plug brass
interface with the cold plate of the dewar vessel containing LN2. The
round 1~inch plug is  attached to the bottom part of the "spider" and
the socket sits on the  stainless steel spring mounted on the cold plate
of the dewar vessel. The  socket is connected with the cold plate with
four copper straps and serves as  the first temperature reduction stage
between the dewar cold plate held at LN2  temperature and chip operational
temperature. When the dewar is closed, the  spring load on the socket
ensures good thermal contact between socket and  plug, secured even more
by a thin layer of thermal conductive cryogenic  grease. Width and
length of both temperature reduction stages copper straps are adjusted in 
such a way that in the open loop the temperature of the chip equilibrate at 
approximately ${-110}$~C. The dewar hold-time is about 28--30 hours depending 
on ambient temperature, thus refilling it once  a day is fully sufficient for 
keeping good thermal stability of the CCD  detector. 

The dewar is vacuumized through a valve attached to the side wall of the
dewar.  The Varian-7000 vacuum pump is the standard equipment of our
observatory and  it takes a few hours to pump out the dewar. Typical
vacuum obtained reaches  ${5\times10^{-6}}$~mbar. The pumping cycle is
repeated every 2--3 months, and  is performed without removing the dewar
from the telescope. 

The dewar is attached to the 16~mm thick aluminum flange of 360~mm
diameter (Fig.~2).  A light aluminum box (called "golden box")
containing CCD electronics is also  attached to that flange. The whole
instrument is then mounted to the  positioning device which allows
rotation of the dewar along the optical axis  and positioning the CCD
detector in respect to RA and Dec. The positioning  device is bolted to
the "instrument mounting plate" (IMP) -- a 16~mm thick  aluminum plate
mounted 225~mm behind the bottom of the telescope mirror  cell. The IMP
is supported with seventeen ${50\times50}$~mm machined aluminum  posts
which minimize flexure of the entire plate. There are two additional 
steel tee-beams which make the plate even more stiff. Test measurements
showed  that flexure is negligible over the entire range of the
telescope zenith  distances as designed. The room between IMP and bottom
of the mirror cell is  used for autoguiding system and large filter
wheel (see Section~6). Also  150~mm diameter electromechanic shutter
(Prontor-Werk, Wildbad, Germany, Model  E-150) is mounted there as well
as all necessary cabling. External covers are  mounted to the edge of
the IMP and supporting posts protecting the instrument  from the
external light and making the instrument compact.  Upgrade with the 
next generation cameras will mean  a simple replacement of the
"external" first  generation camera dewar. 

\vspace*{6pt}
{\it 5.1. CCD Electronics}
\vspace*{6pt}

The electronics for CCD Camera was designed as a modular system to allow easy 
upgrade  for next generation multichip  mosaic cameras. The system was 
designed to  accommodate any of common scientific grade 3-phase CCD imagers 
available on the  market. 

The system  consists of eight printed circuit boards. All boards were
designed  in "hole-through" standard technology. Three small boards are
placed in the  dewar, four boards are plugged to the motherboard mounted
in the  "golden box" close to the dewar. The last board is the ISA
(AT-Bus) interface  card for PC-compatible computers. The analog part of
electronics (preamplifier, clock-driver, bias and signal boards -- see
below) bases on schematics kindly provided by Dr.\ Jim Gunn and designed
by him for the Sloan Sky Digital Survey project. Block diagram of the CCD
electronics and  data acquisition system is shown in Fig.~5. 

Below we give a short description of the boards:

\vskip6pt
{\it 5.1.1 CCD Socket Board}

This is a rectangular board with rectangular hole in the center to allow
the  cold finger of the dewar cryogenic system reaching  the bottom of
the chip  package. Traces connect appropriate clock signals of CCD
sections and by  selection of appropriate jumpers one can select
required output amplifiers of  the chip. All signals go from CCD socket
pins to miniature signal, clocks and  temperature sensor connectors from
which harness of gold plated stainless  steel wires provide electrical
connection with other dewar boards. The CCD  socket board must be
designed separately for each specific CCD type to be  used. 

\vskip6pt
{\it 5.1.2 Dewar Preamplifier/Clock-Driver Boards}

The dewar preamplifier board and clock-driver boards are
${78\times52}$~mm  printed boards piggy-back mounted and attached to the
CCD "spider" mounting in  the dewar. Preamplifier board contains 2
independent channels and each bases  on OPA627 (Burr-Brown) operational
amplifier. Analog inputs are AC-coupled  with polystyrene capacitors.
The clock driver board operates with analog  switches: DG403 and DG445
(Siliconix). Connectors mounted at the top and  bottom of each board
connect the boards with CCD socket board and hermetic  connector mounted
on the side wall of the dewar. 

\vskip6pt
{\it 5.1.3 Motherboard}

The motherboard is a ${265\times120}$~mm rectangular printed board mounted
on  the bottom of the "golden box". It contains a data bus connecting
signal/bias  modules and the CCD controller. The modules are plugged in with
96-pin DIN  connectors. The bus contains 4 slots for signal/bias modules and
one for CPU  controller. Currently, for the first generation CCD camera,
only controller  and one signal/bias slots are used.  The board contains also
power supply filters  and CCD temperature stability circuit driver. 

\newpage
\begin{picture}(200,500)(10,0)
\thicklines
\put(50,0){\framebox(140,50){~}}
\put(60,10){\shortstack{\large DATA ACQUISITION\\
~\\
\large WORKSTATION}}

\put(230,10){\shortstack{\large SUN 10/512\\
\large (OBSERVER'S HOUSE)}}

\put(110,70){\vector(0,1){20}}
\put(110,70){\vector(0,-1){20}}

\put(120,70){\shortstack{\large "DATA NETWORK"}}

\put(50,90){\framebox(120,90){~}}
\put(50,90){\framebox(120,30){~}}
\put(50,90){\dashbox(120,60){~}}
\put(60,96){\shortstack{\large PC ISA\\
\large CONTROLLER\\
~\\
~\\
\large TAXI\\
\large RECEIVER\\
~\\
~\\
~\\
\large FAST ETHERNET\\
100 Mbps}}

\put(50,205){\framebox(120,140){~}}
\put(50,205){\dashbox(120,55){~}}
\put(50,205){\framebox(120,95){~}}
\put(50,205){\dashbox(120,115){~}}
\put(65,212){\shortstack{\large BIAS BOARD\\
~\\
~\\
~\\
\large SIGNAL BOARD\\
~\\
~\\
~\\
\large MASTER\\
\large  CONTROLLER\\
\small DSP TMS320C50\\
~\\
~\\
~\\
\large SECONDARY\\
\large  CONTROLLER\\
\small (OPTIONAL)\\
\small DSP TMS320C50}}

\put(190,135){\line(0,1){110}}
\put(190,135){\vector(-1,0){20}}
\put(190,245){\vector(-1,0){20}}

\put(195,135){\shortstack[c]{T\\ A \\ X\\ I\\~\\~\\ L\\ I\\ N\\ K\\
\\~ \\
125\\
Mbps
}}

\put(220,115){\shortstack{\large CCD PC\\
\large (TELESCOPE BUILDING)}}

\put(240,240){\shortstack{\large GOLDEN BOX}}

\put(50,370){\framebox(120,40){~}}
\put(50,370){\dashbox(120,20){~}}
\put(65,375){\shortstack{\large PREAMPLIFIER\\
~\\
~\\
~\\
\large CLOCK-DRIVER}}

\put(50,445){\framebox(100,30){~}}
\put(80,450){\shortstack{\large CCD \\
~\\
\large SOCKET}}

\put(70,500){\framebox(70,20){~}}
\put(85,505){\shortstack{\large SITe 424B}}

\put(70,500){\line(0,-1){8}}
\put(80,500){\line(0,-1){8}}
\put(90,500){\line(0,-1){8}}
\put(100,500){\line(0,-1){8}}
\put(110,500){\line(0,-1){8}}
\put(120,500){\line(0,-1){8}}
\put(130,500){\line(0,-1){8}}
\put(140,500){\line(0,-1){8}}

\put(250,420){\shortstack{\large DEWAR}}

\put(20,155){\line(0,1){143}}
\put(20,155){\vector(1,0){30}}
\put(20,298){\vector(1,0){30}}

\put(22,160){\shortstack[c]{S\\ E \\ R\\ I\\ A\\ L\\ ~\\~\\ L\\ I\\ N\\ K\\
\\~ \\
6\\
Mbps
}}

\put(30,380){\line(0,1){80}}
\put(30,380){\vector(1,0){20}}
\put(30,400){\vector(1,0){20}}
\put(30,460){\vector(1,0){20}}

\put(190,310){\line(0,1){90}}
\put(190,310){\vector(-1,0){20}}
\put(190,330){\vector(-1,0){20}}
\put(190,380){\vector(-1,0){20}}
\put(190,400){\vector(-1,0){20}}
\end{picture}

\vspace*{.5cm}
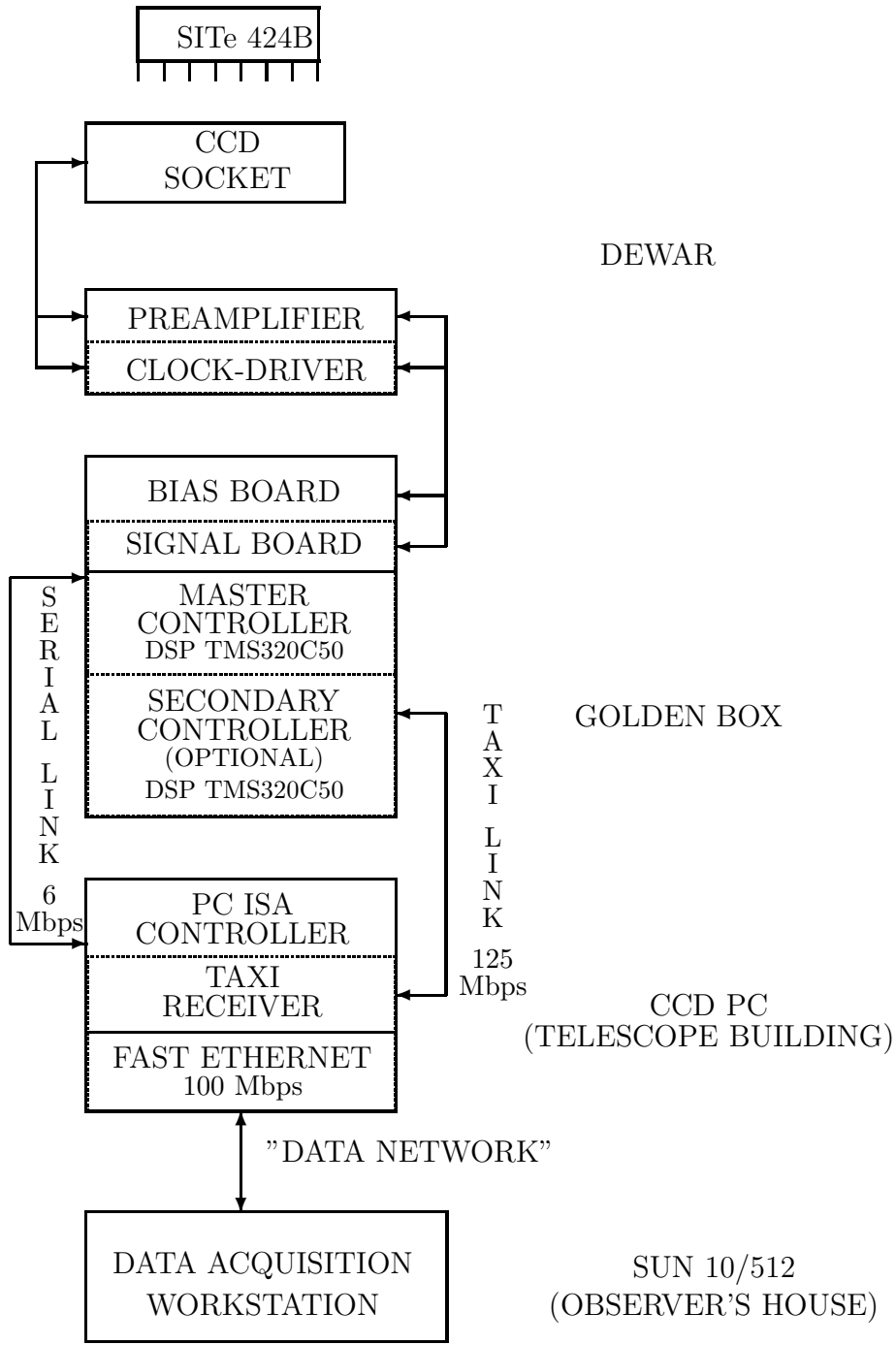
{\footnotesize Fig.~5. Block diagram of the main CCD and data acquisition 
system.}

\newpage
\vskip6pt
{\it 5.1.4 Signal Board}

Signal board is a dual-slope correlated sampling circuit (DSCS) and 16
bit AD  converter. DSCS uses OPA627 (Burr-Brown) operational amplifiers,
DG445  (Siliconix) analog switches and polystyrene integration
capacitor. The ADC is  100~kHz CS5101A (Crystal) converter. Its serial
output is converted to a parallel one with 74HC595 shift registers. The
board size is a standard 3U  EuroCard with 96 pin DIN connector for
plugging it into the motherboard. Two independent channels are
available on the board. 

{\it 5.1.5 Bias Board}

The bias board is a piggy-back mounted card on the signal board. Its
size is  slightly smaller than that of the signal board. It can provide
up to 24  voltage lines. Voltage of each line can be set with  software.
MAX528 (Maxim) DA  converters are used for voltage generation and LT1021
(Linear) serves as  voltage reference. Also the CCD temperature
stabilization circuit is placed on  this board. 44~pin connector at the
top of the board provides connection for  clocks, biases, analog video
and digital signals with the dewar boards {\it via} connectors placed on the
side wall of the "golden box".

\vskip6pt
{\it 5.1.6 Controller Boards}

Master controller board is a standard 3U EuroCard with 96~pin connector
at the  bottom edge. The controller generates all necessary digital
signals for clock  drivers, signal and bias boards and the shutter
control signal. All the  signals are generated by the digital signal
processor (DSP) TMS320C50 (Texas  Instruments) running at 50~MHz. The
processor has 10k 16~bit RAM memory, large enough to run even quite
complicated microcodes. 16k external EPROM memory  serves as a storage
of the {\sc dsp monitor} program as well as CCD software and clocking patterns.
Communication with the external world is done over a synchronous  6~Mbps
serial line. In the case of single CCD chip camera and one amplifier 
reading the same line can serve as pixel transmission line.
Communication up to  several dozen meters with the 50 Ohm coaxial cable
can be achieved with  75121/2 line drivers. Also 8 MHz clock signal for
ADC chips on signal board is generated on the controller board. 

For reading multichannel systems the second, optional board can be piggy-back
mounted on  the master controller card. It consists of the additional
digital signal  processor TMS320C50 and contains also the TAXI
transmission circuit (AMD Am7968 chip) allowing transmission of the
digital data with speed up to 125~Mbps  with the pair of 50~Ohm coaxial
cables over the distance of a few dozen meters. During the normal
operation the processor is in the "idle mode",  dissipating  minimal
amount of energy. It is waked-up for transmission by  the interrupt
signal generated by master DSP from the main signal board.  Similarly to
the main signal board DSP, {\sc monitor} and transmission software are  stored 
in the 16k external EPROM and are loaded to the processor RAM memory after 
reset. Processor can also be programmed from an external computer {\it via} 
master controller board DSP. 

\vskip6pt
{\it 5.1.7 PC ISA (AT-Bus) Controller Board}

The ISA controller board is one of the series of interfaces
between  external computers and the CCD controller DSPs. We decided
to design ISA  interface  first and to use PC-type hardware because of
its low cost, sufficient durability of good quality PC-hardware,
real-time operation  under DOS operating system and easy interfacing.
The PC computer was designed  to serve mostly as data buffer between
DSPs and external workstation. It can,  however, operate also in
standalone mode, and in this mode it is usually used  for system
development. 

The ISA controller contains receiver and transmitter for synchronous
serial  line to communicate with the master controller board DSP over
six coaxial  50~Ohm cables (three for transmission and  three for
receiving data). For  multichannel systems it contains also TAXI
receiver (AMD Am7969) for 125~Mbps TAXI transmission line. Received data from
this interface are stored in computer  memory with DMA transfers. 

The PC computer is a Pentium 133 MHz computer and is stored in the
control room  at the ground level of the telescope dome. It is connected
with  the observer's house  computers with the Fast Ethernet 100~Mbps
network.  In the normal "relay mode" of operation, after collecting the
entire CCD-line  data, the PC sends such a record over the Fast Ethernet
network to the data  acquisition workstation from the observer's house.

\vspace*{6pt}
{\it 5.2. The CCD Camera Software}
\vspace*{6pt}

The CCD system software consists of three parts: DSP microcontrollers 
software, PC-software and data acquisition workstation software (see 
Section~7.2.1). 

The DSP microcode was written in TMS320C50 assembly language. It does
all the  work to control the CCD reading, exposing etc.\ The following
modules reside in  the DSP processor memory of the master controller
board after reset:\\ 
-- {\sc monitor} program. This program initializes and controls DSP after reset. It
allows also to interface DSP with the external computer to perform
interactive operations. Simple commands like memory displaying
and modifying, sending/reading data to/from ports, loading program and
data memory and executing programs are implemented.\\
-- {\sc obs} program. This is the kernel of the CCD control system -- program
executing all functions of the CCD system. One can set all reading
parameters, subraster size, type of reading (normal mode/driftscan),
start/stop exposure, read the chip, wipe the chip etc.\ The program
communicates with the external computer over the synchronous serial line
with a simple protocol sending appropriate commands and statuses of
executed commands.\\
-- {\sc bias} program. Program for programming bias voltages.\\
-- {\sc cpu2} program. Program for interfacing with the optional second controller 
board DSP.\\
-- {\sc pixel data}. Serial clocks pattern for pixel reading.\\
-- {\sc line data}. Parallel clocks pattern for line shift.\\
-- {\sc bias data}. Data for setting appropriate voltages by bias program.

Bias program is executed automatically after reset ensuring proper
voltages of clock signals and biases. Thanks to software approach to
the control of the  system it is very easy to implement new functions,
new clocking patterns etc.\ This makes the system very flexible and
easy to interface with practically any  type of 3-phase CCD detectors. 

The software of 
the optional second controller board DSP for multichannel data reading 
contains of 
the following modules:\\
-- {\sc monitor} program. Similar to above-mentioned  monitor of the master board DSP
for interactive work with the external computer ({\it via} DSP of the master
controller board).\\
-- {\sc obs2} program. The program for sequential reading of the subsequent channels 
and sending data {\it via} TAXI transmitter to the external computer. The processor 
is normally put in the "idle mode" in which it draws only minimum current 
dissipating  minimum energy. It is then awaken by the interrupt from the 
master board DSP and programmed to read and transmit requested number of data. 
The program reads all requested channels and transmits data over the TAXI 
link. When all data are sent it goes again into the "idle mode". Currently 
reading up to 16 channels is implemented, though the system is capable to 
read/send up to 32 channels data during the typical 20~$\mu$s pixel time.\\
-- {\sc channel data} masks. Information which channels should be read.

The PC software was written in C-language  and includes the following
programs:\\
-- {\sc dspmon} program. This program serves for interactive communication
with the master control board DSP monitor.\\
-- {\sc cpu2mon} program. This program serves for interactive communication
with the secondary board DSP monitor.\\
-- {\sc obs} program. The main program for interfacing CCD system with an external 
computer. Executed without parameters, the program works in the local, 
standalone mode with the PC computer as external host. In the normal "relay 
mode", the program opens an Internet socket connection with the external 
workstation. When connected, it works as a relay transmitting protocol 
messages between workstation and CCD controller. Also pixel data obtained line 
by line from the CCD controller are transmitted to the external workstation.\\ 
-- DSP TMS320C50 assembly language development tools. Commercially available 
package of development tools like assembler, linker, hex converter and 
simulator programs.

Currently the single chip, first generation camera with the SITe detector is
read  with one channel only. Though it could be easily read over the
serial 6~Mbps  synchronous line which serves for communication with
PC~ISA interface card, it  is actually read  with the second controller
board {\it via} TAXI link as it  were a multichannel system. Thus, the
future upgrade with next generation  mosaic camera will be a simple
adding of a few additional signal/bias modules  for other chips and some
minor changes in the software. 

The first generation CCD camera performs very well. Table~1 shows the
basic  reading parameters for three reading speeds: fast, medium and
slow (pixel time  15~$\mu$s, 18~$\mu$s and 24~$\mu$s, respectively). 

\begin{center}
T~a~b~l~e~~1\\
Parameters of the first generation CCD camera

\begin{tabular}{|l|c|c|c|}
\hline                    
Reading Mode              &       SLOW  &   MEDIUM   &   FAST\\       
\hline                    
Pixel time ($\mu$sec)     &        24   &    18      &    15 \\              
Gain ($e^-$/ADU)           &       3.8   &   7.1      &   9.9 \\
Readout Noise ($e^-$)     &       5.2   &   6.3      &   7.4 \\
Reading time (sec)        &       110   &    82      &    69 \\
\hline
\end{tabular}
\end{center}

\vspace*{6pt}
{\bf 6. Autoguiding System and Filter Wheel}
\vspace*{3pt}

Fig.~4 presents the autoguider and filter wheel before attaching to the
telescope.  The autoguiding system of the 1.3-m Warsaw telescope is a
fully automatic, computer controlled device. It consists of the probe
with a small prism  reflecting the off-axis beam toward the guider CCD
detector. The probe is  mounted on x-y stage and can move over the large
part of the entire 1.5~degree  field of view of the telescope.
Additionally the guider CCD camera can move  toward the probe prism for
focusing. All three motions are realized with the very precise lead, ball
screws manufactured by Warner Electric GmbH (Wolfschlugen,  Germany) and
stepper motors. Accuracy of positioning in ${x-y}$ coordinates is 
better than 0.03~mm over the entire accessible range.  There is an area
in   which the probe obscures the main CCD. It is, however, protected
and  unaccessible for users by software. 

The initial positions in the ${x,y}$ and focus coordinates are set with 
infrared LED and phototransistor sensors. The LED beam is cut by a
sharp blade  and transition of the corresponding phototransistor signal
defines the zero  position. Repeatability of zero position is better
than 0.02~mm. All motions  are secured by limit switches and by
software. The stepper motors are  interfaced with the computer using
commercial translator drives (type  SS2000MD4, Warner Electric, Bristol,
Ct, USA). 

The autoguider is a separate module which is bolted to the instrument
mounting  plate (IMP). To minimize length of cables the translator
drives are mounted also on the IMP as close as  possible to the motors
driven by them. The  electronics driving the autoguider is placed in the
red anodized aluminum box  called "red box" mounted on the side wall
between the IMP and the bottom of  the mirror cell. 

The block diagram of the autoguider, filter wheel and data
acquisition system is shown in Fig.~6. The autoguider detector is a
EEV37-10 CCD chip (EEV, Chelmsford, UK). This is a $512\times 512$, 15~$\mu$m 
pixels backside illuminated detector. It is a frame transfer device  very 
well suited for autoguiding purposes. It is driven by almost identical 
electronics as the main CCD camera. Only the bias  and clock driver boards 
were slightly modified to accommodate positive voltages of high and low  rails 
of chip clocks. The chip is mounted in a small aluminum head and is  
thermoelectrically cooled. The head is filled with dry nitrogen and  
hermetically sealed. The chip is cooled to approximately ${-25}$~C with  
temperature stabilized by electronic circuit on the bias board. The heat  
dissipated by the thermoelectric cooler is negligible at this temperature. 

The preamplifier and clock driver boards are mounted outside the CCD
head but  as close as possible. The signals go from these boards to the
bias/signal  board plugged to the small motherboard mounted in the "red
box". This  motherboard is designed for one signal/bias board and controller
card only but it also includes some line drivers for autoguider and filter
wheel motors and control  system sensors. 

\newpage
\begin{picture}(250,530)(0,-40)
\thicklines
\put(50,0){\framebox(140,50){~}}
\put(60,10){\shortstack{\large DATA ACQUISITION\\
~\\
\large WORKSTATION}}

\put(230,10){\shortstack{\large SUN 10/512\\
\large (OBSERVER'S HOUSE)}}

\put(110,70){\vector(0,1){20}}
\put(110,70){\vector(0,-1){20}}

\put(120,70){\shortstack{\large "DATA NETWORK"}}

\put(50,90){\framebox(120,60){~}}
\put(50,90){\framebox(120,30){~}}
\put(60,96){\shortstack{\large PC ISA\\
\large CONTROLLER\\
~\\
~\\
~\\
\large FAST ETHERNET\\
100 Mbps}}

\put(50,225){\framebox(120,90){~}}
\put(50,225){\framebox(120,40){~}}
\put(50,225){\dashbox(120,63){~}}
\put(65,232){\shortstack{\large BIAS BOARD\\
~\\
~\\
~\\
~\\
\large SIGNAL BOARD\\
~\\
~\\
~\\
\large GUIDER\\
\large  CONTROLLER\\
\small DSP TMS320C50
}}

\put(250,470){\framebox(15,12){~}}
\put(290,470){\framebox(15,12){~}}
\put(330,470){\framebox(15,12){~}}
\put(250,466){\line(0,1){20}}
\put(290,466){\line(0,1){20}}
\put(330,466){\line(0,1){20}}
\put(250,476){\line(-1,0){5}}
\put(290,476){\line(-1,0){5}}
\put(330,476){\line(-1,0){5}}

\put(230,380){\framebox(120,70){~}}
\put(250,392){\shortstack{\large AUTOGUIDER\\
\large MOTORS\\
\small AND\\
\large  POSITION\\
\large SENSORS
}}

\put(230,225){\framebox(120,70){~}}
\put(250,232){\shortstack{\large FILTER WHEEL\\
\large MOTOR\\
\small AND\\
\large  POSITION\\
\large SENSORS
}}

\put(250,310){\framebox(15,12){~}}
\put(250,306){\line(0,1){20}}
\put(250,316){\line(-1,0){5}}

\put(170,240){\line(1,0){40}}
\put(190,240){\vector(-1,0){1}}
\put(210,240){\line(0,1){200}}
\put(210,280){\vector(1,0){20}}
\put(210,440){\vector(1,0){20}}

\put(220,115){\shortstack{\large GUIDER PC\\
\large (TELESCOPE BUILDING)}}

\put(50,350){\framebox(120,40){~}}
\put(50,350){\dashbox(120,20){~}}
\put(70,355){\shortstack{\large PREAMPLIFIER\\
~\\
~\\
~\\
\large CLOCK-DRIVER}}

\put(50,425){\framebox(100,30){~}}
\put(80,430){\shortstack{\large CCD \\
~\\
\large SOCKET}}

\put(-20,400){\shortstack{\large GUIDER\\
\large PROBE}}

\put(-20,270){\shortstack{\large RED\\
\large BOX}}

\put(70,480){\framebox(50,20){~}}
\put(72,485){\shortstack{\small EEV37-10}}

\put(70,480){\line(0,-1){5}}
\put(80,480){\line(0,-1){5}}
\put(90,480){\line(0,-1){5}}
\put(100,480){\line(0,-1){5}}
\put(110,480){\line(0,-1){5}}
\put(120,480){\line(0,-1){5}}

\put(20,128){\line(0,1){133}}
\put(20,128){\vector(1,0){30}}
\put(20,261){\vector(1,0){30}}

\put(23,133){\shortstack[c]{\small S\\\small E \\\small R\\\small I\\\small A\\ 
\small L\\ ~\\~\\\small  L\\\small  I\\\small  N\\\small  K\\
\\~ \\
\small 6\\
\small Mbps
}}

\put(30,360){\line(0,1){80}}
\put(30,360){\vector(1,0){20}}
\put(30,380){\vector(1,0){20}}
\put(30,440){\vector(1,0){20}}

\put(190,280){\line(0,1){100}}
\put(190,280){\vector(-1,0){20}}
\put(190,300){\vector(-1,0){20}}
\put(190,360){\vector(-1,0){20}}
\put(190,380){\vector(-1,0){20}}

\put(0,-20){\shortstack{
{\footnotesize Fig.~6. Block diagram of the autoguider, filter wheel and data 
acquisition system.}}}
\end{picture}

The controller board with DSP TMS320C50 processor is very similar to that of 
the main CCD camera master  controller. Additionally it serves also as 
autoguider and filter wheel controller. It is connected with the external  
computer with the 6~Mbps synchronous serial line (six 50~Ohm cables). The  
line serves also for CCD pixel data transmission. The external computer is a 
PC with 120~MHz Cyrix 586 processor. Similar to the main camera PC it is  
connected to the Fast Ethernet "data network" and in the normal "relay mode"  
operation serves as a relay between the observer's house  data acquisition  
workstation and the guider DSP controller. 

The autoguider controller software is also very similar to that of the main CCD 
camera. The main module -- {\sc gui} -- drives the guider CCD camera in a
similar way  as {\sc obs} drives the main system. Additionally it includes
procedures for running the  stepper motors of the autoguider
(simultaneously in three axes) and filter  wheel, checking position of
filters and guider etc. Remaining modules are also  adjusted for the EEV CCD
detector. 

The PC software includes procedures which calculate centroid of the guide star  
images and derive necessary telescope motion corrections. The corrections are  
then applied to the telescope control system by four miniature relays mounted 
on the  PC ISA controller board which are driven by four lines of the output 
port mounted on  that board. 

Extensive tests show that the guider performs according to
specifications.  Although its full frame field is relatively small --
$2.2\times 2.2$ arcmins --  guiding stars can be easily found with the
"find star" procedure implemented  in the Telescope Control System data
acquisition software (see Section~7.2.1) which selects  stars
appropriate for guiding from the HST Guide Star Catalog and calculates 
required probe coordinates. The range of accessible field of view is so
large  that even in very empty fields the user has a list of a few dozen
stars  suitable for guiding. The guider performs well with stars up to
13 mag with  1~s exposure. For guiding a subwindow $30\times 30$ pixels
centered on the  guiding star is read. The reading time of the guider
CCD is so short that such a subraster can be read up to 30 times per
second.  Thus, the autoguider is well suited for future implementation of some 
seeing correction instruments like tip-tilt secondary mirror etc.\

The filter wheel is a classic design of disk mounted off the optical
axis.  To accommodate large filters for the next generation multichip
camera the  filters up to 160~mm diameter can be mounted in the 8 slots.
Filters can be up  to 8~mm thick. Due to large filter size, the disk
diameter is 690~mm. The disk is rotated by a stepper motor with two
stage gears and friction gear made with  a roller and the outer edge of the
disk. The gear ratios are selected in such a way that the full rotation
of the disk takes approximately 20~s. The  position of the filters is
controlled by  three pairs of infrared LED --  phototransistor sensors
mounted below and above the disk. The hole pattern in  the disk codes the
proper position of the filter and number of the filter  slot. Accuracy
of positioning is about 1~arcmin. Although the filter wheel  assembly is
very stiff, the position of the filter can be double-checked at  any
moment with the sensor. 

Both, filter wheel stepper motor and position sensor, are controlled by
the  DSP TMS320C50 microprocessor of the controller board in the "red box".
The  procedures for moving filters, checking position of the filters are
included  in the {\sc gui} program module. Slots 1--5 are reserved for {\it
UBVRI} filters.  Glass Schott filter combos are used for approximation
of the standard bands.  Filters are anti-reflective coated with ${\rm
MgF_2}$. First observations show  that except for the ultraviolet filter
which short wavelength limit is reduced  by transmittance of the field
corrector the remaining filters are very close to  the standard system. 

\vspace*{9pt}
{\bf 7. Observer's House Equipment}

\vspace*{6pt}
{\it 7.1.Hardware}
\vspace*{6pt}

The observer's house equipment consists of a data acquisition computer, data 
reduction computers, data storage disks, data archiving tape drives and 
networking hardware. 

The data acquisition computer is a two processor Sun Sparc 10/512
workstation  with 192~MB memory running Solaris 2.5.1 operating system.
The computer  has two additional Fast Ethernet cards installed. This
machine serves as the main  data acquisition host. It is connected with
the dome PC "relay computers"  with the Fast Ethernet 100~Mbps "data
network". The second Fast Ethernet  controller is connected to the Fast
Ethernet "computer network" -- independent  network connecting data
acquisition workstation with all data reduction hosts  in the house. The
standard Ethernet port of the data acquisition computer is  connected to
the local network of the Las Campanas Observatory and further  with the
Internet. Thus, the data acquisition computer serves as router  between
all these networks. It is also connected {\it via} the RS232 serial link with
the telescope control PC located in the control room in the  telescope
building. 

Beside running the data acquisition software, the data acquisition
computer  runs also data pipeline software for automatic flat-fielding of
the collected frames. 

Currently there are three computers which are used as data reduction
hosts: two processor Sparc Ultra 2200 with 192~MB memory running 
Solaris 2.5.1 and two double Pentium-Pro 200MHz computers with 64~MB
memory  running RedHat Linux~4.2. All computers are connected to the
Fast Ethernet  "computer network". Pentium-Pro hosts are used as "number
crunching boxes"  with no monitor and keyboard and are accessible only
{\it via} network. All three  computers run the OGLE-2 data reduction pipeline
software and are capable to  perform photometric reductions and data
analysis of all collected during the night images within 24
hours. 

Due to the huge data flow, the data storage system must have large
capacity to  ensure easy and comfortable work. Currently about 90~GB of
disk storage is  available. 26~GB of storage disks are connected to the
data acquisition  workstation for collected raw data. Remaining disks
are connected to the Ultra  2200 workstation for storage of reduced
data. All disks are accessible from any  host over the network with no
degradation in performance due to sufficient  throughput of the Fast
Ethernet "computer network". 

To archive huge amount of collected and reduced data a good, reliable,
high  capacity medium is necessary. The DLT tape subsystem was selected
as primary  data archiving hardware. Good reliability, fast data
transfer and fast data  access proved  the DLT drive to be a good
choice. Typical capacity of a DLT  tape is about 25~GB. Two DLT drives
are available. Also standard, 2~GB Exabyte  8200 drive is available as
a backup system. 

\vspace*{6pt}
{\it 7.2. Software}
\vspace*{6pt}

The software packages running on data acquisition and reduction computers  
include the following parts: data acquisition software, data pipeline software 
for automatic data  flat-fielding and OGLE-2 photometry data pipeline 
software. Short description of packages is given below. 

\vskip6pt
{\it 7.2.1 Data Acquisition Program}

The data acquisition software consists of the set of server programs and
Telescope Control System {\sc (tcs)} program providing the observer with a
graphical
user interface. The servers are  responsible for communication with and
control of devices they serve. On the  other hand they communicate with
the {\sc tcs} program letting the latter to  execute many different tasks
simultaneously. As communication between servers and {\sc tcs} program is
established with  Internet socket protocol, the {\sc tcs} program can be run
on any host in the Internet network making it possible to  control the
system remotely from any place in the world. At present the  following
servers are implemented:\\
-- {\sc tcs\_server}. This server  controls the communication between the
telescope and {\sc tcs} program.\\
-- {\sc ccd\_server}. This server  establishes an Internet socket protocol
connection  with the dome PC controlling the CCD camera and controls the
main CCD  camera, receives and writes to disk pixel data (FITS format is
the standard) and communicates {\it via} "shared memory" with an external image
display which  displays automatically collected CCD frames.\\
-- {\sc gui\_server}. This program  establishes an Internet socket protocol 
connection with the dome PC controlling the autoguider and filter wheel. This 
server controls guider motion, guider CCD operation and filter wheel 
operation.\\ 
-- {\sc gsc\_server}. This is a local server interfacing the {\sc tcs} program 
with the HST Guide Stars Catalog (selection of guiding stars, etc.). 

The {\sc tcs} program is a X-windows based graphic user interface for easy 
and comfortable work. It consists of three main windows for controlling
the  telescope, CCD camera and guider, respectively. 

The telescope window is the main window which opens when the program is 
invoked. The user can start and stop telescope tracking, slew the
telescope to  the selected object, set the focus, turn on/off dome and
slit tracking of the  telescope or park the telescope at zenith
position. Tracking rates in both RA  and DEC can be set or changed,
software hand paddle can be used for small  movement of the telescope.
For better notification of the end of current activity, voice
messages are implemented. The object to which the telescope is  to be
slewed can be selected in a few ways. First, simply its coordinates can 
be given. Secondly, offsets from the current position can be specified. 
Alternatively, the object can be selected from libraries of objects: the
 standard telescope library including many interesting objects in the
sky, or from custom library prepared by the user. Finally, each pointing
of the telescope is  remembered and the object can be selected from
session history. The popular public domain graphic program  {\sc xephem} is
also interfaced with the {\sc tcs} software, and by clicking the  mouse on
requested star in the {\sc xephem} SkyView Map the user can select it as  a
slewing target. 

Two windows, which can be opened from the instrument menu, control  CCD
camera and guider. The CCD camera window allows to start/stop  the
exposure, select exposure time, number of exposures, type of object and 
object name (if the object is selected from the library its name is set 
automatically). Also some reading parameters including  size of the
subraster to be read,  hardware binning, readout speed, gain, mode of
reading can be selected. There  are two modes available:  normal mode,
that is standard exposure/reading  sequence, called also "still frame"
mode, and driftscan mode where the
chip is continuously read with the rate  synchronized with the telescope
tracking. With the present setup driftscans in  declination are
possible: the telescope tracks in RA with sidereal rate and 
additionally tracks in declination. The CCD detector is adjusted in such
a way  that its columns are parallel to meridians. Reading rate of the
chip is  synchronized with the declination tracking rate of the
telescope. Resulting  image is a strip with the width of the chip and
length equal to declination  tracking rate multiplied by driftscan time
(practically only computer file  size of the image is the limitation
here). Scanning along declination (that is  the great circle which
meridian is) allows to obtain driftscan images at any  declination
making the entire sky available to that technique. Classic  driftscans
in RA are usually limited to the vicinity of the equator. 

The CCD window also contains a submenu from which many standard
procedures  can be executed. These include focus exposure (set of
exposures on one frame  with different foci and the telescope shifted
slightly between them; the best  focus can be easily selected by
inspecting stellar images) and automatic collection of  calibration
images in all modes (dome, sky and driftscan flat-field images as  well
as biases can be obtained). These automatic procedures reduce 
considerably burden on the observers. 

Guider window allows to move the probe over the telescope field of view,
focus  it, make full field image for guide star selection, setting the
exposure time  and other reading parameters and  start/stop guiding. The
automatic procedure  for selection of guide star is also implemented. It
allows to select  appropriate stars based on the HST Guide Star Catalog.
When executed, the user  gets the list of appropriate stars and after
selection of the best star the  probe is ready for moving to the
calculated coordinates. Starting of guiding  is as simple as making full
frame image, clicking on the displayed guide star  with the mouse and
clicking the start button. The ${30\times30}$ pixels subframe  with guide star
and reference cross is continuously displayed in the image  display part
of the window. Corrections are displayed in the information part of  the
guider window. The user can also display a plot of corrections {\it vs.}
time. 

The {\sc tcs} software can be run in two modes. The normal mode is the 
standard mode when the user interactively operates the telescope and 
instruments. The second mode, called batch mode allows to execute the 
telescope-language batch scripts, which contain sequence of telescope, 
CCD and guider commands. When such a batch is started the telescope and
instruments execute  the subsequent commands allowing to operate in
fully automatic, preprogrammed  robotic mode. Majority of system
functions are available in batch mode, as well as some  simple script
command  like loops, variables, labels etc., making this mode a  very
powerful way of observing. First, the number of human mistakes is
minimized.  Secondly, the efficiency of the system is much better than
during manual operation.  The batch mode is the main mode of OGLE-2
regular observations. 

The last element of the data acquisition software is the image
display. This  is a simple and easy to use image display which
integrates the best features  of publicly available software of this
type. It is interfaced with the {\sc ccd\_server} program {\it via} "shared
memory" mechanism and displays automatically all frames taken by the CCD
camera. Simple data analyzing tools like profile and contour  
displaying, FWHM fit, calculation of aperture magnitude, zooming and
flipping  the image as well as printing are implemented. 

\vskip6pt
{\it 7.2.2  Data-Pipeline}

The standard data pipeline consists of the software for automatic
preparation  of the flat-field images and automatic flat-fielding of the
data immediately  after they are collected. The software is prepared for
both types of collected  frames - normal mode and driftscans. 

Full set of flat-field images is kept for both modes of observations.
When new  flat-field images are collected they must be  preprocessed first
with automatic  software and then compared with "old" flat-fields. The
user can decide whether to  replace the "old" flat-field images. For driftscan
mode the one dimensional flat-field and bias images are prepared by
averaging the flat-field and bias images along columns  with a special
procedure to remove stars which are sometimes present on the flat-field
images.  Normal mode flat-field and bias frames are preprocessed with the
standard IRAF\footnote{IRAF is distributed by National Optical Observatories, 
which is operated by the Association of Universities for Research in 
Astronomy, Inc., under cooperative agreement with National Science 
Foundation.} routines from {\sc ccdproc} package. 

The automatic flat-fielding is performed by daemon programs:
{\sc flatdaemon} for normal mode images and {\sc driftdaemon} for
driftscan mode images. Both  programs wait for incoming images,
recognize their type and perform  appropriate bias and flat-field
corrections. The normal mode flat-fielding  bases on IRAF {\sc ccdred}
package software while the driftscan flat-fielding is  done by a special
software written for that purpose. Logs of calibrated frames 
including all basic information like date, time, object, filter, air mass,  
fwhm and sky level  are also prepared by both daemons. 

The flat-fielding software ends the standard data-pipeline giving the
user  calibrated data, free from instrumental effects. Behind this point
a custom  made software for further reductions can be installed similar
to the OGLE-2 photometry software which is described in the next
subsection.

\vskip6pt
{\it 7.2.3  OGLE Photometry Data-Pipeline}

The OGLE photometry  data pipeline consists of a set of daemon programs
which wait for  flat-fielded FITS images and start photometric reduction
software on collected, calibrated frames. To use the CPUs most effectively, 
one daemon is started for each CPU.  Daemons have protection mechanism against 
running reduction of the same frame  more than once. 

Since the OGLE-2 data are collected mostly in the driftscan mode the
reduction  software is tuned to this type of data. Because the driftscan
images show substantially variable PSF due to variable  seeing in
different moments of the scan,  inaccuracies in telescope tracking etc.,
the original strip of ${2048\times8192}$  pixel is divided to 64
${512\times512}$ pixels subframes in which the PSF can  be treated as
constant in the first approximation. Photometric reduction is done with the
modified version of {\sc dophot} photometric software (Schechter, Saha and
Mateo 1993). The {\sc dophot} program is run in fixed position mode --
positions of the stars are  fixed when the photometry is derived using
the list of positions from the very good seeing template image. This makes
reductions much faster and also much more reliable in the case of frames
taken at poor seeing conditions. Additionally the {\sc dophot} is run in 
variable PSF mode to account for small PSF changes sometimes still
present on  the subframes. 

The template for a given field is prepared from reduction of the best
seeing  frame. A special procedure runs reduction of subframe
enlarged by 50~pixels in all possible directions
and ties  results to one photometric system based on photometry of
overlapping areas  (100~pixels). When making the final template
list the subframe is trimmed to the original size and
overlapping regions are removed.
Thus, there is no multiple photometered stars in
the final list. The {\it I}-band  images are treated as
"master-templates" -- list of detected {\it I}-band  stars is used as
input list  when preparing templates for other colors. In  this way the
position of a given star in the final list is the same in all  colors,
making further data handling considerably easier. 

When reducing non-template frames, first the shift between the frame and
the  template image is derived with cross-correlating the small
subframes from the  central part of both -- current and template
-- driftscan strips. Then, the  frame is divided into subframes, taking into
account just determined shift and  photometry of each subframe is derived.
Finally, photometry of each subframe is  tied to the photometric system
of the template. 

For easy manipulating of huge amount of photometric data, databases
containing   photometric results for each field and color are
constructed. Software  described by Szyma{\'n}ski and Udalski (1993)
with some modifications is used.  Additional modules for on-line data
analyzing can be installed at this point.  One of such modules -- the
Early Warning System (EWS) for on-line detection of  non-variable
objects changing their brightness will be implemented when the
appropriate  sample of photometric measurements of observed fields is
collected. 

Additional package of programs of OGLE-2 software consists of set of
programs  for data handling and archiving. Those include programs for
easy and safe data  archiving on magnetic tapes.

\vspace*{12pt}
{\bf 8. Performance of the System. Prospects.}
\vspace*{9pt}

The OGLE-2 project started regular observations with the telescope and
instrumentation system  on January 6, 1997. The first target was Large
Magellanic Cloud. 20 fields  covering more than 4.2 square degrees of the
LMC were selected. Later on  additional targets were added: 0.7
square degree in Galactic disk regions in  Carina, 10 square degrees in
the Galactic bulge, and 2.3 square degrees in the  central parts of the
Small Magellanic Cloud. All observations are made in  driftscan mode
with the effective exposure time 120~s, 180~s and 240~s for  {\it I, V}
and {\it B} filters, respectively for Magellanic Clouds and Carina 
fields and 85~s, 128~s and 170~s for Galactic bulge fields. Majority of 
observations are made with {\it I}-band filter, with some additional images in 
{\it  V} and {\it B} colors.

The main advantage of the driftscan mode used for OGLE-2 observations is
the large field which can be covered with the ${2048\times2048}$ pixel
detector.  High  sensitivity of new generation CCD detectors allowing
short exposures with  relatively deep range makes such a mode ideal for
survey of dense stellar fields  where the crowding of stellar images is
the limiting factor. The main drawback  of driftscan method is somewhat
lower seeing resolution because of small  non-uniformities of telescope
tracking. It is, however, still very good with  the median value about
1.1~arcsec. In the normal mode images with seeing as  good as
0.75~arcsec have already been collected in the {\it I}-band, though  the
sample of the normal mode images is not very large. 

The driftscan OGLE-2 images are limited to 8192 lines for easy data
handling.  Thus, the size of the frame is ${2048\times8192}$ pixels,
(34~MB of raw data including prescan).
During the  night it is possible to collect between 50 (long
exposure) and 70 (short exposure)  frames giving the data flow of the
order of 1.7--2.3~GB/night. A total data flow  of raw data is estimated
to be up to 600~GB/year. The data flow is about 30  times larger than
during the OGLE-1 part of the OGLE project. 

Table~2 presents the main current targets of the OGLE-2 project. Average
 number of stars detected is also given there. One can notice that the
number  of observed stars is about 30 millions for bulge fields and
about 7 millions in LMC. As some fields are observed a few times per
night up to 45 millions of  photometric measurements per night are
obtained for dense bulge fields, and  about 15 millions of measurements per
night for LMC. All these measurements are  derived and analyzed in the
near-real time -- within 24 hours. 

\begin{center}
T~a~b~l~e~~2\\
{Main targets of the OGLE-2 project (season 1997).}

\begin{tabular}{|l|c|c|c|c|}
\hline
TARGET                 &  LMC     &  Galactic Disk &  Galactic Bulge & SMC \\  
\hline
Field area (sq.\ deg). &   4.2    &   0.7          &   10            & 2.3 \\
Number of stars        &  $7\times10^6$  & $0.6\times10^6$ & $30\times10^6$ & $2\times10^6$\\
\hline
\end{tabular}
\end{center}

Fig.~7 shows a tiny sample of variable stars detected in LMC fields to present 
quality of data. Light curves include observing points collected during the 
first six weeks of observations only. Figs.~8 and 9 show similar sample of 
periodic and long term variables from the Galactic bulge, respectively. The 
absolute calibration of these {\it I}-band light curves is preliminary --
accurate to 0.1~mag only.  Finally, all photometric data should be calibrated 
with accuracy of 0.01--0.02~mag. Finally, Fig.~10 presents the first 
microlensing event found in the OGLE-2 data -- OGLE-BUL-20. This is an event 
97-BLG-56 detected earlier by the MACHO group (Becker {\it et al.} 1997), and it was 
used for testing our EWS software which will be fully implemented after the 
first season of observations of each field. 

The current setup of the telescope will be used until the second
generation  camera is available. It will be a mosaic of at least eight ${2048
\times4096}$  pixel chips with 15~$\mu$m pixels. With such a new equipment it will
be possible  to monitor almost an order of magnitude larger areas,
reaching more than  $10^8$ measurements per night. It is expected to
start that new phase of the  project sometime around~1999. 

Continuing the tradition of the first part of the OGLE project most of 
collected data -- photometry of microlensing events, variable stars,
catalogs  of stars in all observed fields etc.\ will be available for the
astronomical  community over the Internet network from OGLE archive:
{\it sirius.astrouw.edu.pl} or its mirror {\it astro.princeton.edu.} 

\Acknow{It is our pleasant duty to acknowledge the most valuable help and 
friendly  support received from many individuals and Institutions throughout 
the development of our project. 

We are particularly grateful to Professor Bohdan Paczy\'nski for his continuous, 
enthusiastic interest in the project and zillions of helpful suggestions. 

Our most sincere thanks are due to the Carnegie Institution of Washington with 
its President Dr.\ Maxine Singer and to the Observatories of the Carnegie 
Institution of Washington with their Directors Dr.\ Leonard Searle and Dr.\ 
Augustus Oemler for generous granting of the telescope site at the Las 
Campanas Observatory what was a crucial moment in the history of the  project. 
We are also grateful to Dr.\ Ian Thompson for his friendly support at all 
stages of the project. 

Our warm thanks go to the Staff of the Las Campanas Observatory and in 
particular to its Director Dr.\ Miguel Roth for his bearing with constantly 
good humor and friendliness the additional burden of our presence on the 
Mountain. It is also our pleasant duty to acknowledge the most valuable help 
received in many different occasions during both construction and operation 
phases of our telescope from our tried friends Wojtek Krzemi\'nski, Bill
Kunkel, Hernan Solis, Oscar Duhalde, 
Emilio Cerda, Frank Perez, Peter de Jonge, Juan Luis Lopez, David Trigo 
and Patricio Pinto. 

Our thanks are also due to Princeton University for financial and moral 
support, particularly at early phases of the microlensing project when its 
success was not evident. In particular we are indebted to Dr.\ Jim Gunn for 
letting us using his design of analog electronics for Sloan Digital Sky 
Survey and his facilities and labs at Princeton University Observatory as 
well as many helpful comments and suggestions. We also thank George Pauls and  
Brian Elms for their help during construction of the CCD camera. 

We appreciate very much financing of the project by the State Committee for 
Scientific Research through the grants Nos.\ 2~1173~91~01, 538/IB/115/93, 
1142/IA/115/95 to MK and 2P03D02908 to AU. Generous financial help was 
received in a difficult moment from the Foundation for Polish Science. Partial 
support for this project was provided with the NSF grant AST-9530478 to 
B.~Paczy\'nski. Continuous financial and organizational support from the 
Copernicus Foundation for Polish Science is also thankfully acknowledged.} 

\newenvironment{references}%
{
\footnotesize \frenchspacing
\renewcommand{\thesection}{}
\renewcommand{\in}{{\rm in }}
\renewcommand{\AA}{Astron.\ Astrophys.}
\newcommand{\AAS}{Astron.~Astrophys.~Suppl.~Ser.}
\newcommand{\ApJ}{Astrophys.\ J.}
\newcommand{\ApJS}{Astrophys.\ J.~Suppl.~Ser.}
\newcommand{\ApJL}{Astrophys.\ J.~Letters}
\newcommand{\AJ}{Astron.\ J.}
\newcommand{\IBVS}{IBVS}
\newcommand{\PASP}{P.A.S.P.}
\newcommand{\Acta}{Acta Astron.}
\newcommand{\MNRAS}{MNRAS}
\renewcommand{\and}{{\rm and }}
\section{{\rm REFERENCES}}
\sloppy \hyphenpenalty10000
\begin{list}{}{\leftmargin1cm\listparindent-1cm
\itemindent\listparindent\parsep0pt\itemsep0pt}}%
{\end{list}\vspace{2mm}}

\def\TYLDA{~}
\newlength{\DW}
\settowidth{\DW}{0}
\newcommand{\dw}{\hspace{\DW}}

\newcommand{\refitem}[5]{\item[]{#1} #2%
\def\REFARG{#3}\ifx\REFARG\TYLDA\else, {\it#3}\fi
\def\REFARG{#4}\ifx\REFARG\TYLDA\else, {\bf#4}\fi
\def\REFARG{#5}\ifx\REFARG\TYLDA\else, {#5}\fi.}

\newpage
FIGURE CAPTIONS

Fig.~1. 1.3-m Warsaw telescope facilities at Las Campanas Observatory. 

Fig.~2. The 1.3-m telescope and instruments. 

Fig.~3. "First generation" camera dewar with CCD mounting.

Fig.~4. Autoguider module and filter wheel mounted on Instrument Mounting 
Plate before attaching to the telescope. 

Fig.~7. Sample of periodic variable stars detected in LMC after six
weeks of observations. Numbers below each light curve indicate period in
days. Magnitude scale is accurate to 0.1 mag.

Fig.~8. Sample of periodic variable stars from the Galactic bulge.
Numbers below each light curve indicate period in
days. Magnitude scale is accurate to 0.1~mag.

Fig.~9. Sample of long term variable stars from  the Galactic bulge.
Abscissa is JD hel.: 2450550 -- 2450690. Magnitude scale is accurate to 
0.1~mag.

Fig.~10. Microlensing event OGLE-BUL-20.
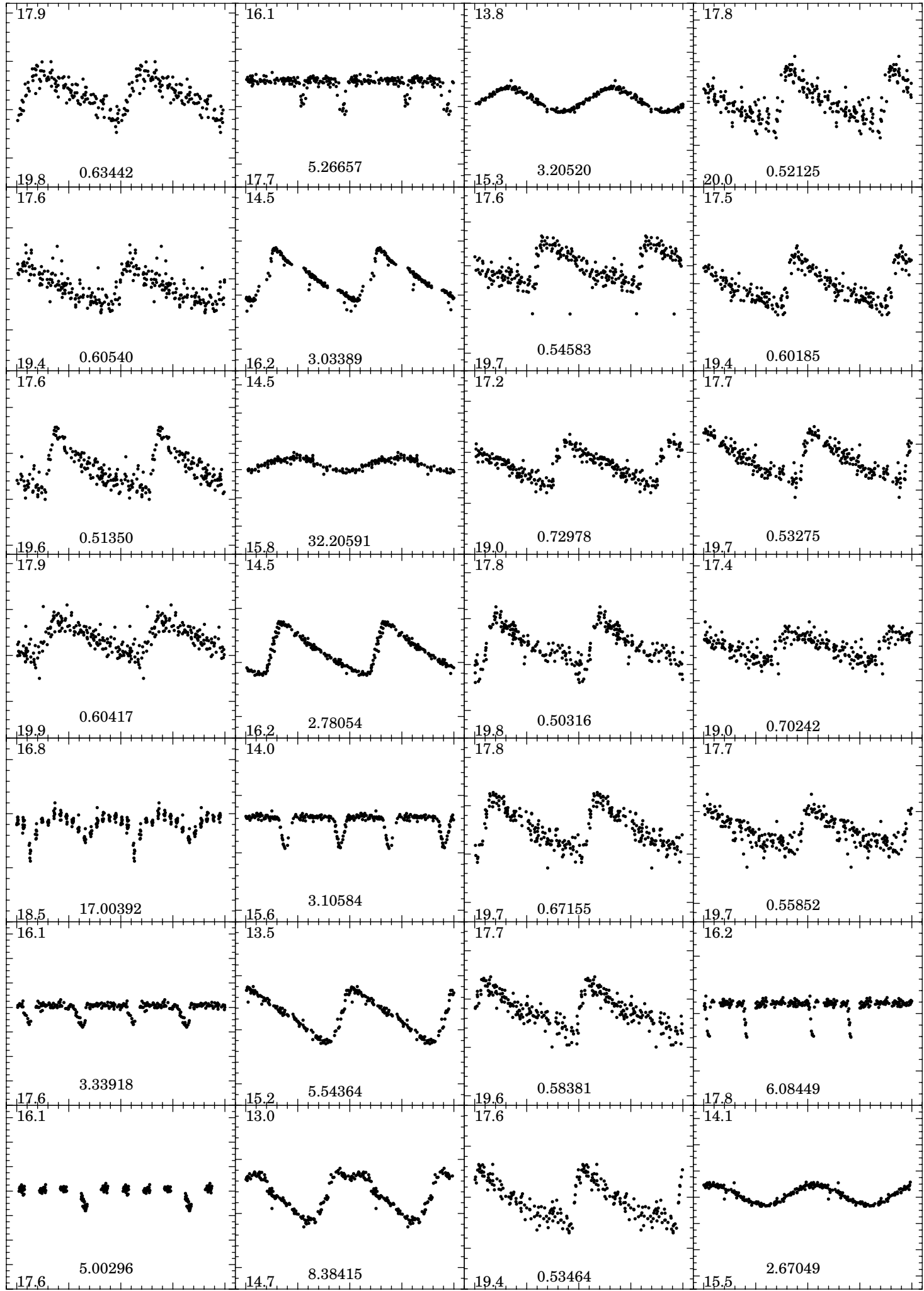
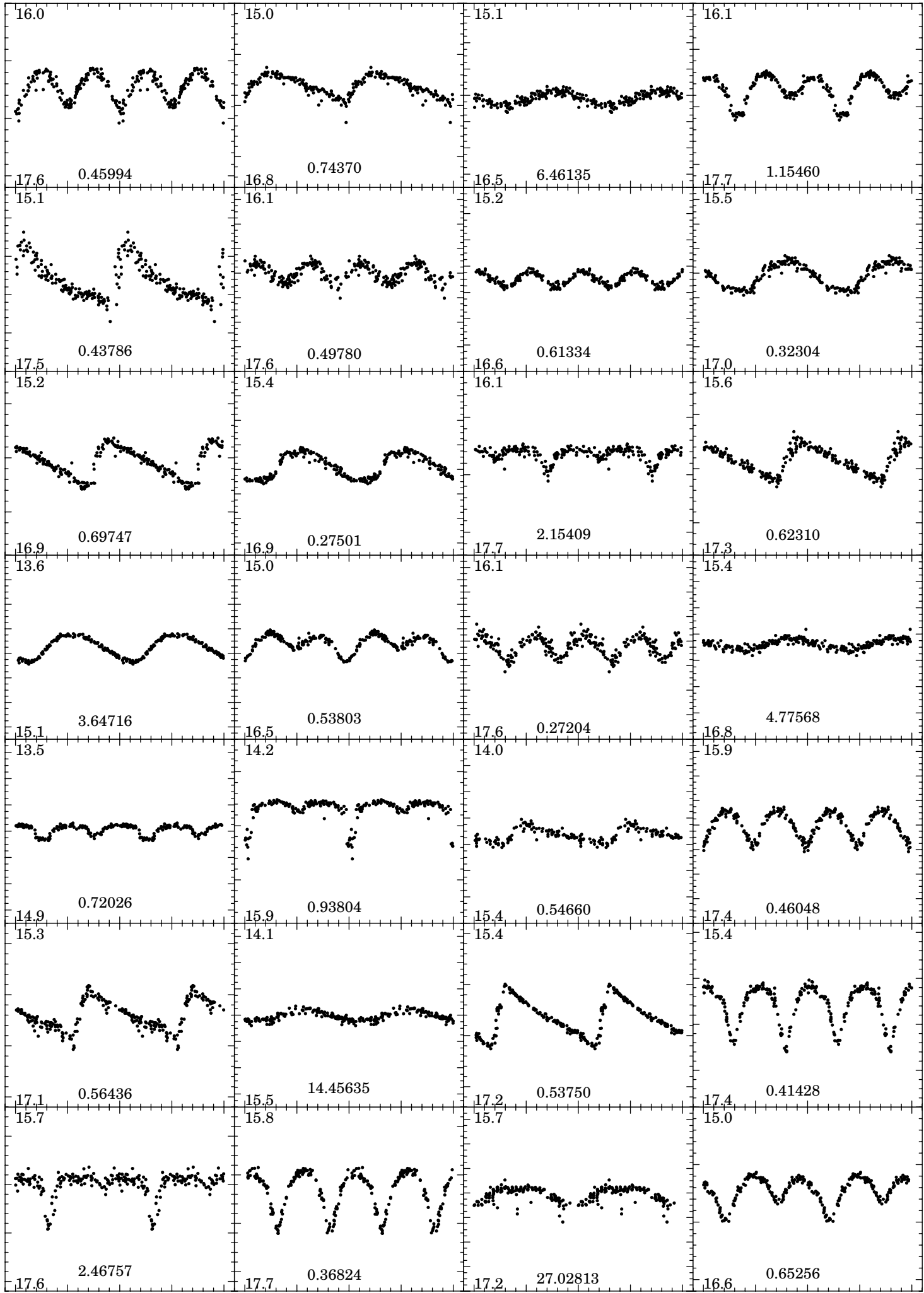
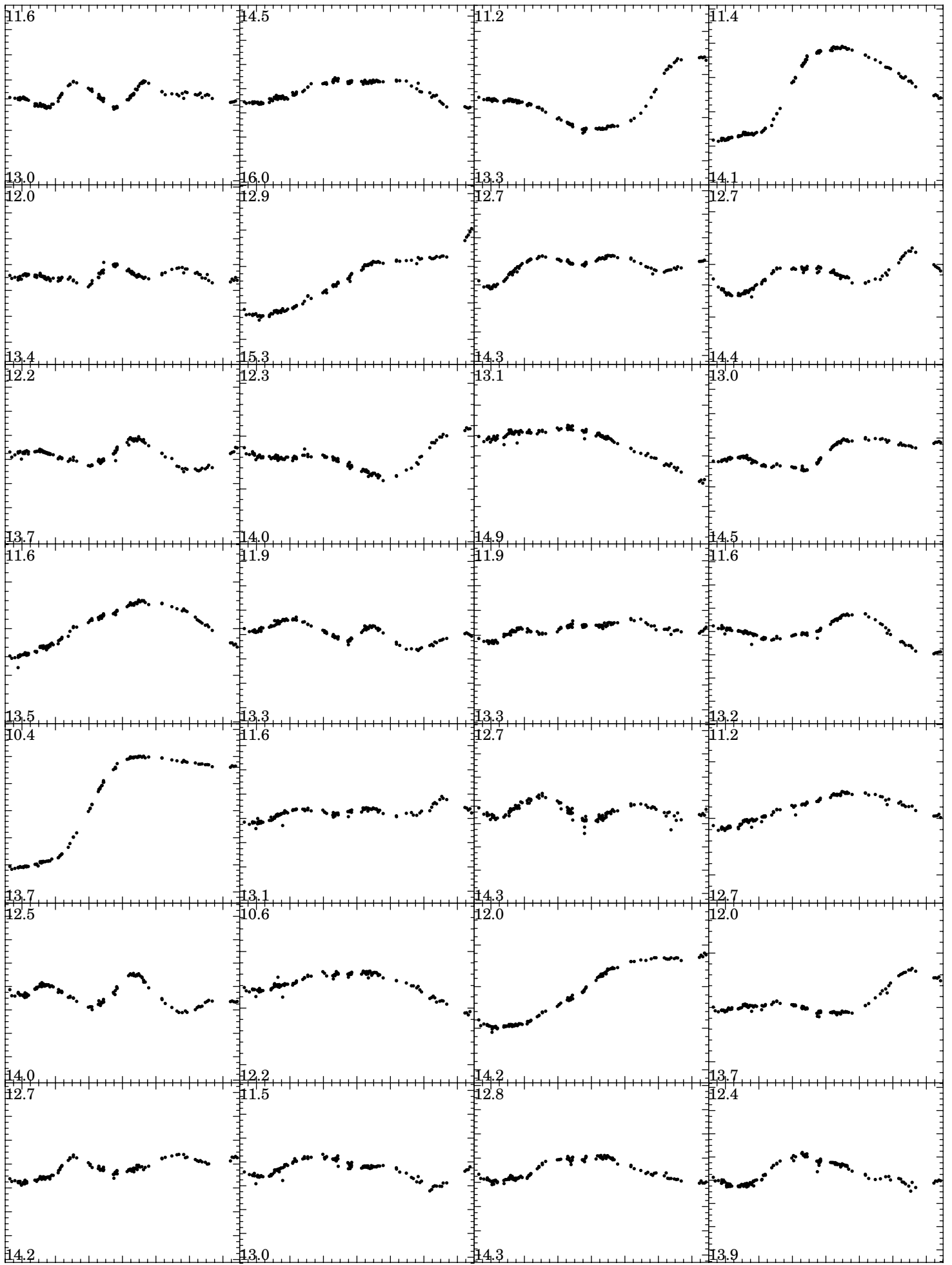
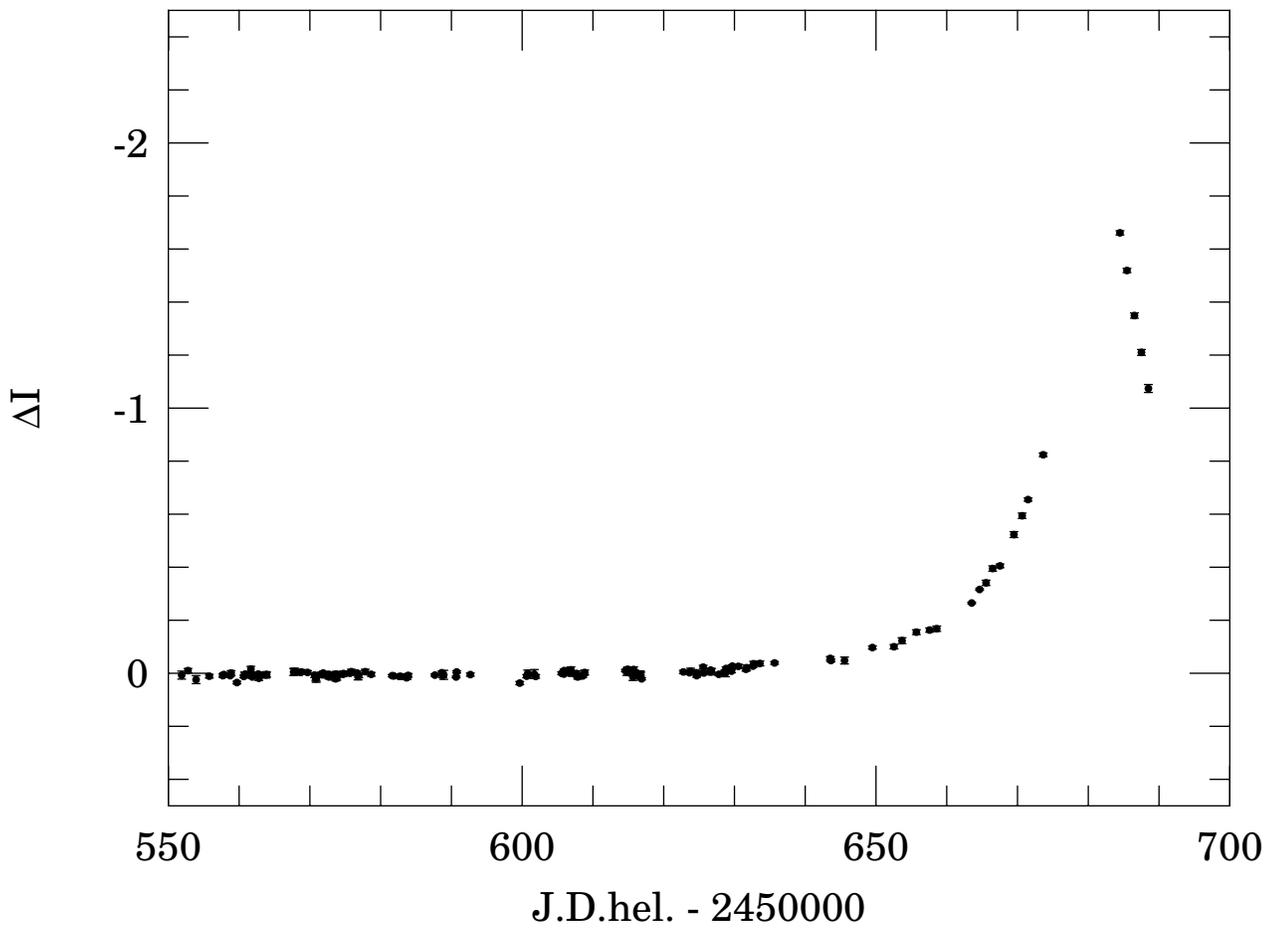
\end{document}